\documentclass[%
 reprint,
 amsmath,amssymb,
 aps,
]{revtex4-2}

\usepackage{graphicx}
\usepackage{dcolumn}
\usepackage{bm}
\usepackage{url}

\begin{document}


\title{A search for a cosmologically-relevant boson in muon decay}

\author{J.I. Collar}
\email{collar@uchicago.edu}
\affiliation{%
Enrico Fermi Institute, Kavli Institute for Cosmological Physics, and Department of Physics\\
University of Chicago, Chicago, Illinois 60637, USA
}%


\date{\today}

\begin{abstract}
Experiments looking for a lepton flavor-violating decay $\mu^{+}\!\!\rightarrow \!e^{+} X^{0}$ are reviewed in light of present-day germanium detector technology, with an eye on scenarios where a long-lived, slow-moving massive boson $X^{0}$ might have a cosmological impact. A broad  swath of interesting, unexplored parameter space very close to the kinematic limit of the decay is found to be within the reach of a new proposed search.  A number of possible roles for $X^{0}$ in past and present  epochs can be investigated.   
\end{abstract}

\maketitle


The study of the decays and properties of the muon holds a prominent place in the development and validation of the Standard Model (SM) of particle interactions \cite{mot1,mot2}. Being the lightest unstable particle in the SM, but still massive enough to be of interest at $m_{\mu}=$105.6 MeV, it is expected to decay exclusively into electrons, photons and neutrinos, and to do so through the weak force, avoiding the  theoretical uncertainties of strong interactions. Its single known decay mode ($\mu\rightarrow e \bar{\nu_{e}} \nu_{\mu}$, plus radiative variations such as $\mu\rightarrow e \bar{\nu_{e}} \nu_{\mu} \gamma$) has a long lifetime of $\tau_{\mu}=2.2 ~\mu$s. This facilitates precision measurements of its properties in intense muon beams, while making them available over a broad range of energies. Recent possible departures from the SM in the muon sector (g-2 anomalous magnetic moment  \cite{g21,g22}, flavor-changing B$^{0}$ decays \cite{lhcb}) have generated renewed attention to this area.

Lepton flavor violation has been experimentally established via the observation of neutrino oscillations, evidencing the incompleteness of the SM. Charged lepton flavor violation (CLFV) is guaranteed to appear by this neutral-particle precedent, in some instances with good prospects for observation \cite{clfv2}. As such, CLFV has been fervently searched for in new modes of muon decay, namely $\mu\rightarrow e \gamma$, $\mu\rightarrow 3e$, and $\mu N\rightarrow e N$ (muon-to-electron conversion in the field of a nucleus) \cite{clfv1,clfv2}. The best limits have been set using intense muon beams available at the Paul Scherrer Institute, probing branching ratios (BR) down to $\sim 10^{-12}$. These searches favor the use of antimuons, due to their larger yield in proton collisions, and in order to avoid the complications associated to muon capture backgrounds \cite{clfv1}. Future upgrades to these searches \cite{clfv3,clfv4} hold promise of casting light on the origin of the g-2 anomaly \cite{g23}.

\begin{figure}[!htbp]
\includegraphics[width=.9 \linewidth]{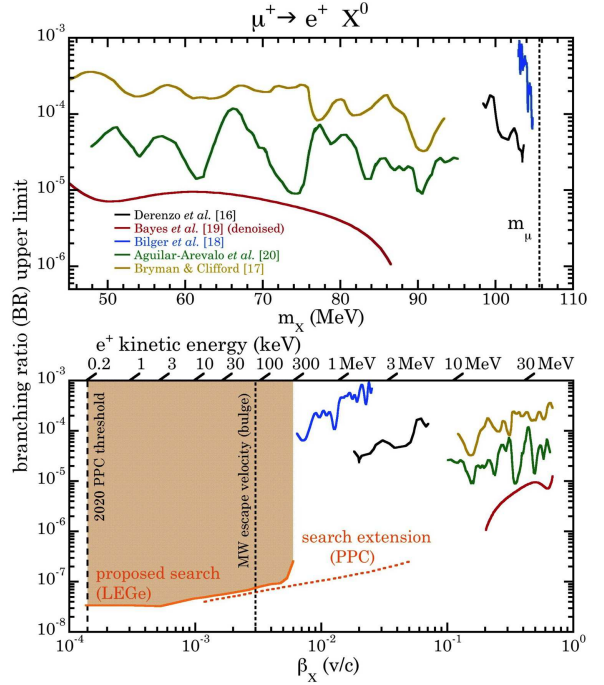}
\caption{\label{fig:epsart} {\it Top:}  conventional display of  sensitivity to \mbox{$\mu^{+}\rightarrow e^{+} X^{0}~$} in BR vs.\ $m_{X}$ space. A vertical line indicates the  mass of the parent muon.  {\it Bottom:} Alternative view  in BR vs.\  speed of emitted boson. Vertical lines denote the escape velocity from the Milky Way, and the lowest energy  measurable by modern point-contact germanium detectors \cite{ESS,cdex}. }
\end{figure}

Numerous extensions of the SM involving symmetries broken at the $>\!\!>\!1$ TeV scale predict hypothetical new particles with lepton-flavor violating couplings, lighter than the muon, and in some instances able to explain standing anomalies \cite{carlos1,carlos2}. Examples are many: axions, axion-like particles, majorons, familons, light gauge Z' bosons, dark photons, etc. (\cite{mot3}, and references therein). In these models, the exotic muon decays  mentioned above are usually heavily suppressed, necessitating a switch in focus to the $\mu\rightarrow e X^{0}$ channel \cite{mot3}, where $X^{0}$ is a new neutral boson.  From an experimental point of view, this possibility has been explored in the two-body decay $\mu^{+}\rightarrow e^{+} X^{0}$, by studying the (Michel) positron energy spectrum of decays from conventional  $\mu^{+}\rightarrow e^{+} \bar{\nu_{e}} \nu_{\mu}$, inspecting it for a superimposed anomalous monochromatic peak determined in its position by the mass of the new boson, $m_{X}$ (see \cite{search1,search2,search3,search4,search5}, with a massless boson search at the spectral endpoint in \cite{search6}). The sensitivity of these searches is restricted to BR $\gtrsim 10^{-5}$, due to the limited energy resolution of the large calorimeters employed, and to the background imposed by ever-present Michel positrons. The search proposed here bypasses both limitations, reaching down to BR $\gtrsim 10^{-8}$ in an unexplored  region of phase space, of possible cosmological interest. 

Among these searches for $\mu^{+}\rightarrow e^{+} X^{0}$, one by Bilger {\it et al.} (\cite{search3}, Fig.\ 1) deserves special attention. It was performed as an {\it ad hoc} test of a proposed solution to the short-lived KARMEN anomaly \cite{karmen} invoking a $m_{X}=103.9$ MeV boson, put forward in  \cite{sergei} before a rapid resolution was reached through less exciting means \cite{karmensolution}. This search employed a germanium diode doubling as muon-stopping target and positron detector, exploring the positron kinetic energy range 0.3 - 2.2 MeV (102.9 MeV $< m_{X} <$ 104.8 MeV). In this approach, the detector acts as a beam dump, with a muon telescope providing the trigger for registration of two  signals in rapid succession (muon and positron energy depositions in Ge), separated by a time interval in correspondence to $\tau_{\mu}$. The BR sensitivity of the search was limited by a number of hardware issues, including a severe degradation of the energy resolution to $\sim$100 keV full-width-at-half-maximum (FWHM), from the O(1) keV FWHM expected of a germanium detector at the energies of interest. The search did however exclude the possibility, of several discussed in \cite{sergei}, of $X^{0}$ being a scalar particle  -always within the context of the KARMEN anomaly-. It also brought this type of search very close to the kinematic limit where $m_{\mu}$ is fully invested into $m_{X}$ and the positron mass $m_{e}$,  with both having no significant kinetic energy at emission. 

The top panel in Fig.\ 1 shows the commonly used representation of $\mu^{+}\!\rightarrow e^{+} X^{0}$ sensitivity in BR vs.\ $m_{X}$ phase space \cite{search1,search2,search3,search4,search5}. In view of it, one might be tempted to devise new ways to improve on BR, but it would be hard to justify pushing $m_{X}$ any closer to the kinematic limit, based on this portrayal alone. An alternative visualization of the present experimental situation can be reached using the  kinematics  of two-body decay at rest \cite{search5}: $E_{e}=(m^{2}_{\mu}+m^{2}_{e}-m^{2}_{X})/2m_{\mu}$, where $E_{e}=m_{\mu}-E_{X}$ is the total positron energy, and $E_{X}=\gamma ~m_{X}$ is the total boson energy, with $\gamma$ its Lorentz factor. 

This unconventional vantage point (BR vs.\ the speed of the emitted new boson, $\beta_{X}$) is shown in the bottom panel of Fig.\ 1. Its top horizontal axis additionally displays the energy deposition of the second signal ($e^{+}$ kinetic energy) expected in a muon-stopping target/positron detector hybrid like that used by Bilger {\it et al.}, when made small enough to allow for most 511 keV gammas (positron annihilation products) to escape undetected, but large enough to contain the full positron ionization track within. Such practical considerations are further discussed below. 

To establish a liaison with cosmological concerns, a superimposed vertical line shows the escape velocity from the Milky Way $\beta_{MW}$, extrapolated to the central bulge of the galaxy \cite{mw} ($\beta_{MW}\sim 2\times 10^{-3}$ at Earth). A sufficiently long-lived $X^{0}$ -plausible in a number of models \cite{sergei}- emitted with  $\beta_{X}\!<\!\beta_{MW}$ would remain trapped in closed orbits within the deep gravitational potential well of the galaxy, able to contribute to several cosmological scenarios of interest. Evidently, short-lived bosons produced in early epochs with any $\beta_{X}$ can decay into lighter stable dark matter candidates, a possibility often considered in the literature. Conversely, they can be redshifted into a present-epoch cold dark matter candidate   ($\beta_{CDM}<2\times10^{-7}$ \cite{cdm}), if their lifetime $\tau_{X}$  allows for it. However, it is worth remembering that a continuous muon -and perchance $X^{0}$- production takes place not just during stellar collapse \cite{sn}, but also in all stars during episodes energetic enough to generate their pion precursors, e.g., in atmospheric flares \cite{flares}, or following cosmic-ray impacts \cite{cr}. Suitable positron-emitting modes of decay such as $X^{0} \rightarrow e^{+} e^{-} \bar{\nu} \nu$ or $X^{0} \rightarrow e^{+} e^{-} \phi$ (where $\phi$ is a massive boson stable or eventually decaying into $\bar{\nu}\nu$ \cite{sergei}), together with the right combinations of BR, $\beta_{X}<\beta_{MW}$, and $\tau_{X}$ can lead to the buildup of a galaxy-bound or star-bound $X^{0}$ population able to contribute a solution to the long-standing mystery of the 511 keV gamma emission from the Milky Way bulge \cite{5111,5112}. Characteristics of this emission hard to accommodate using conventional astrophysical sources, such as its spherical symmetry around the bulge, or the need for modest (few MeV) positron injection energies \cite{john,john2} are explainable with a long-lived MW-bound $X^{0}$ in the $m_{X} \simeq m_{\mu}$ mass range, if decaying as above. Separately, it is worth mentioning that  dark matter solutions to the 511 keV emission riddle favor candidates precisely in the few MeV to few hundred MeV  range \cite{511dm1,511dm2,511dm3}. 

A more detailed study of the  possibilities hinted at above is beyond the scope of this brief note. However, it seems reasonable to expect that suitable regions in BR, $\beta_{X}$, $\tau_{X}$ phase space exist for which $X^{0}$ can play cosmological roles, and that different production and survival/decay scenarios should lead to predictions testable by new experiments aiming to probe the ``room at the bottom" made evident by the BR vs.\ $\beta_{X}$ representation of Fig.\ 1. The rest of this work focuses on assessing the phase space that is presently within reach.

From a technological point of view, germanium detectors have undergone a significant evolution since their use by Bilger {\it et al.} in 1999, one that is very timely for the revival of the technique that is proposed here. Specifically, ultra-low noise p-type point contact (PPC) detectors \cite{ppc} now allow to detect sub-keV energy depositions in large germanium crystals. As such, PPCs have found numerous new applications in searches for neutrinoless double-beta decay \cite{majorana,gerda,legend}, dark matter \cite{cogent,cdex}, and coherent elastic neutrino-nucleus scattering (CE$\nu$NS) \cite{ESS,conus1,texono,nugen}. A second vertical line in Fig.\ 1 marks the energy threshold of contemporary PPCs, putting this capability in the context of what is now a reachable $\beta_{X}$ domain. N-type point contact detectors  (\cite{luke}, here denoted as NPCs) share the same reduced noise and low threshold, but are limited in size due to a sub-optimal charge collection, which results in a severely degraded energy resolution for diodes larger than a few cm$^{3}$ \cite{ppc,luke}. Separately, the inert electrical contact thickness at the entry point of a muon beam varies from the $\sim$1 mm lithium-diffused depth in a PPC, to a sub-$\mu$m boron-implanted layer in NPCs. As discussed below, all these considerations impact the choice of modern Ge detector in a renewed $X^{0}$ search, depending on the $\beta_{X}$ range intended for study.

Experimental design constraints have been studied with the help of MCNPX simulations \cite{mcnpx}, taking as point of departure the M13 ``surface" antimuon beamline at TRIUMF, one of several at this facility suitable for this exploration by virtue of their beam purity, flux, and momentum, selectable in the range 20 MeV/c $< p_{\mu} <$ 200 MeV/c \cite{triumf,beamnote}. The simulations include detailed aspects such as the abatement down to a negligible level of beam-related backgrounds from use of a  15 cm-thick, 4 mm-diameter tapered Pb collimator, and muon energy loss in intermediary materials: a  beryllium entrance window to the detector cryostat, and ultra-thin 25 $\mu$m plastic scintillator \cite{eljen} telescope paddles used to provide a trigger to the data acquisition, coincident with a muon stopping in Ge. Specifically, and for the purpose of estimating the reachable sensitivity, a five-day beam  exposure with a conservative \cite{search3} 1,000 Hz singles rate at the germanium detector are assumed in what follows. 

\begin{figure}[!htbp]
\includegraphics[width=.8 \linewidth]{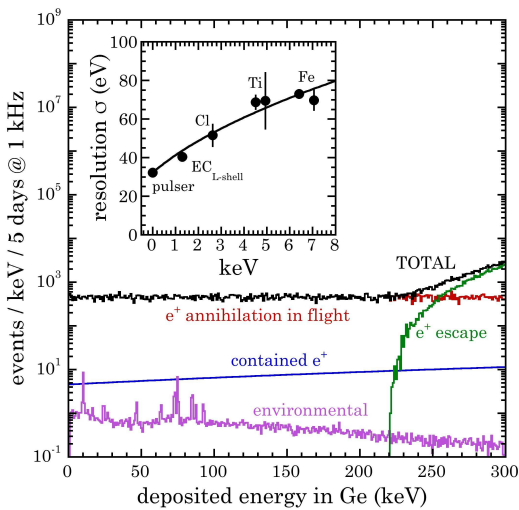}
\caption{\label{fig:epsart} Expected background in the LEGe detector, dominated by annihilation-in-flight and escape of high-energy Michel positrons. A secondary component arises from low-energy Michel positron trajectories  contained within the crystal. Subtractable environmental backgrounds are small by comparison. Lines from Pb fluorescence and $^{68,71}$Ge electron capture (EC) can be identified. {\it Inset:} excellent LEGe energy resolution, measured by the author  using  alpha-induced x-ray emission from labeled materials. }
\end{figure}

Environmental backgrounds able to penetrate the envisioned 5-15 cm-thick Pb shield around the detector were measured rather than simulated, using a 1 cm$^{3}$ GL0110 LEGe detector -a type of NPC \cite{lege}- in the author's laboratory (Fig.\ 2). Besides the mentioned progress in germanium detector technology, other significant advances over the effort at \cite{search3} are planned. For instance, a fast 1 GS/s 16-bit digitization of the Ge detector preamplifier trace over a $20 ~\mu$s window centered around the telescope trigger will allow for the subtraction of steady-state environmental and beam-related backgrounds, and for the identification of sub-keV positron signals even very close in time ($\lesssim1 ~\mu$s) to the initial muon-stopping pulse (the charge collection time of a small LEGe is $\sim50$ ns). Fig.\ 3 illustrates the advantages of performing offline analysis on digitized preamplifier traces, in contrast to the use in \cite{search3} of exclusively post-trigger analog-shaped signals, subject to pile-up and loss of information.

This planned characterization of anti-coincident backgrounds during the $10~ \mu$s pre-trigger period is an approach similar to that implemented during the first CE$\nu$NS detection \cite{science}. While random coincidences with steady-state backgrounds during the 10 $\mu$s post-trigger period are expected to be subdominant vis-\`a-vis the Michel continuum, their removal from the Ge energy spectrum  will eliminate peaks that might otherwise confound the search (Fig.\ 2).  To be considered possible indications of a new boson, events belonging to any remaining peaks must respect the  $\tau_{\mu}$ distribution in their delay from preceding muon-stopping signals.

\begin{figure}[!htbp]
\includegraphics[width=.8 \linewidth]{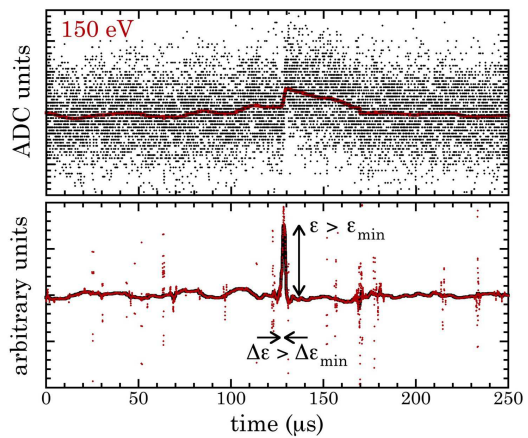}
\caption{\label{fig:epsart} {\it Top:} offline analysis of a digitized LEGe preamplifier waveform containing a 150 eV pulser-induced signal. A line shows the wavelet-denoised trace, displaying characteristic rise- and decay-times. {\it Bottom:}  dots correspond to the fast derivative of the denoised trace. A line joins them following median-filtering. Conditions on minimum amplitude and width for the resulting peak reject low-frequency noise,  pinpointing the onset of {\it bona fide} low-amplitude signals. }
\end{figure}

The lowest testable value of $\beta_{X}$ is determined by the threshold of the germanium diode (Fig.\ 1). In order to push sensitivity in that direction, the best detector electronic noise is necessary. This also has the effect of improving energy resolution (Fig.\ 2, inset), and with it BR sensitivity. Pulse-reset preamplifiers are mandatory towards this goal, and commonplace for PPCs, as they are known to introduce lower noise than resistor-feedback alternatives \cite{goulding,ppc}. A disadvantage however is the maximum signal energy that can be accepted before a reset occurs, which distorts the preamplifier output for O(10) $\mu$s. This  time span is obviously prohibitive for this search, when the initial muon-stopping signal causes the reset. For purposes of the present sensitivity estimate, a reset range corresponding to $\sim$7 MeV (involving a modest increase over typical commercial units) is adopted. This imposes a few-MeV maximum muon energy deposition in Ge, leaving room for positron energy registration.

A beam energy of 4.0 MeV ($p_{\mu}=$ 29.3 MeV), resulting in muon energy depositions of $\sim$3.7 MeV after losses in intermediary materials  is an ideal compromise, as M13 beam purity improves drastically for $p_{\mu}<$ 29.8 MeV \cite{search6}. The simulated depth of muon implantation (i.e., its decay site) in Ge at 3.7 MeV is $330\pm25 ~\mu$m (Fig.\ 4, inset), which  imposes the NPC detector design option due to the inert  surface layer considerations mentioned above. In turn, and similarly to \cite{search3}, this implantation depth dictates the largest $\beta_{X}$ that can be explored, as positrons with kinetic energy $\gtrsim$ 300 keV will not deposit their full energy in Ge if backward-emitted (i.e., opposite to beam direction, Fig.\ 4). For these choices, neither muon nor positron track lengths require a large detector, which respects the maximum size limitation of NPCs. A GL0055 LEGe,  just 8 mm in diameter and 5 mm in axial length \cite{lege} (one fourth the volume of the unit used for environmental background studies, rendering those conservative) is seen as the best commercial germanium diode choice for a $X^{0}$ search reaching the lowest possible values of $\beta_{X}$. For this specific beam configuration, detector geometry, and energy range of interest, the  expected dominant background arises from partial energy depositions by the small fraction of Michel positrons undergoing annihilation with electrons while still in flight \cite{aif1,aif2,aif3}, and those backward-emitted that escape the detector. Their contributions surpass those from low-energy Michel positrons having trajectories fully-contained in Ge (Fig.\ 2).

\begin{figure}[!htbp]
\includegraphics[width=.8 \linewidth]{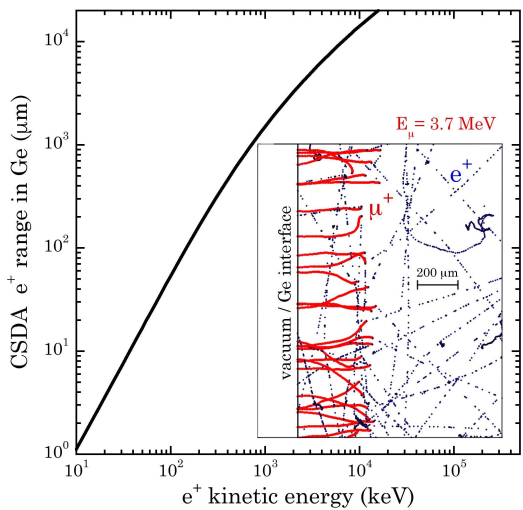}
\caption{\label{fig:epsart} Continuous slow-down approximation (CSDA) positron track length in germanium, derived from \cite{csda1,csda2}. {\it Inset:} MCNPX-simulated depth of antimuon implantation for the LEGe detector run considered in the text. Notice scale. }
\end{figure}

To obtain the BR boundary of the estimated sensitivity region shown in Fig.\ 1, the  total background in Fig.\ 2 is scanned, taking a two-sigma statistical fluctuation in an energy window corresponding to the local FWHM to represent a 95\% C.L.\ excess, and comparing this number of events to the total expected for the run. The FWHM is derived from the energy resolution (Fig.\ 2, inset), using its standard energy dependence \cite{eres1,eres2}. Minor penalties for signal loss to preamplifier resets, and for the simulated probability ($\sim$89 \%) of double 511 keV gamma escape are considered, and applied. Two main factors contribute to the excellent sensitivity foreseen: an improvement by more than three orders of magnitude in detector energy resolution with respect to previous searches using large scintillator calorimeters \cite{search5}, and the modest background rate expected.  A tiny detector size, and the  small fraction of Michel decays at low $E_{e}$ ($\sim\!\!10^{-5}$ for $E_{e}\!<\!1$ MeV) synergistically combine towards this last advantage.

The exploration of faster-moving, lower-mass $X^{0}$ bosons, perhaps part of scenarios where redshift and/or decay following their production in early cosmological epochs contributes a solution to the dark matter problem, would require a larger detector able to contain higher-energy positron tracks (Fig.\ 1). Unfortunately, radiative losses rapidly lead to incomplete energy depositions for positrons above few MeV, even in the largest viable (few kg \cite{ESS,david}) germanium detector, smearing the monochromatic signature sought. Keeping in mind that the critical energy for positrons in germanium is 17.6 MeV \cite{ec}, and in the absence yet of a dedicated simulation, it is possible to estimate that the  ``germanium beam dump" method is then limited to $\beta_{X}\lesssim 0.05$ (Fig.\ 1). 

The larger detector size needed to contain O(1) cm tracks corresponding to few-MeV positrons (Fig.\ 4) necessarily leads to a PPC choice for this search extension. The maximum possible axial length for non-coaxial PPCs  is $\sim$4 cm \cite{ourieee}, even in the presence of a strong gradient of charged impurities in the semiconductor material, co-adjutant to charge collection \cite{ppc}. A muon implantation depth in Ge of 2 cm  can be obtained with a M13 beam energy of 37 MeV (95.8 MeV/c). In a 4 cm-diameter, 4 cm-long PPC, this depth obviates a $\sim$1 mm dead surface layer, and maximizes positron track containment for polarized muons, characteristic of surface beams \cite{mupol1,search6}. A modified (reduced gain) resistor-feedback preamplifier with an enhanced energy range out to 50 MeV is necessary for this search expansion, resulting in an energy threshold penalty. Expected  sensitivity, calculated as above, is shown in Fig.\ 1 (a small improvement in sensitivity for intermediate $\beta_{X}$ arises from the absence of resets in this preamplifier).  In this larger crystal of higher efficiency for gamma detection, a $X^{0}$ spectral signature might have associated satellite peaks at the energies corresponding to an additional single or double 511 keV gamma absorption. This can be exploited in an analysis strategy leading to an enhanced sensitivity.  

New particles awaiting discovery can hide at the kinematic limit, where the products of a decay carry minuscule, barely detectable kinetic energies. If stable or long-lived, such particles are liable to play intriguing cosmological roles, by virtue of their non-relativistic speeds. The search proposed here incorporates three distinct realms: an ultra-high energy regime of new symmetry-breaking, cosmic scales, and the faintest energies  reachable only by state-of-the-art radiation detectors. These are attractive, perhaps sufficient grounds to undertake this initiative. However, a more pragmatic justification exists: considering the numerous particle models that generate a viable $X^{0}$, and the certainty that CLFV must occur in nature, no stone should be left unturned in probing \mbox{$\mu^{+}\rightarrow e^{+} X^{0}~$} parameter space.

The author is indebted to John Beacom, Doug Bryman, Peter Cooper, Dan Hooper, Rocky Kolb, Wilhelm Mueller, and Carlos Wagner. This work was supported by NSF awards PHY-1806722, PHY-1812702, and by the Kavli Institute for Cosmological Physics at the University of Chicago through an endowment from the Kavli Foundation and its founder Fred Kavli.

\bibliography{apssamp}

\begin{thebibliography}{63}
\bibitem [{bea()}]{beamnote}%
  \BibitemOpen
  \href@noop {} {}\bibinfo {note} {The M13 beamline at TRIUMF has been recently
  decommissioned. Other locally available beams such as M20, M15 and M9A are
  suitable substitutes for this search, offering similar flux, energy, and
  purity characteristics.}\BibitemShut {Stop}%
\end{thebibliography}

\end{document}